\documentclass[a4paper, 12pt]{article}
\usepackage[T1]{fontenc}
\usepackage{graphicx,epsf,amssymb,amsbsy,amsfonts,amssymb,amsmath,physics,courier,IEEEtrantools,jheppub,romannum,ytableau}

\def\be{\begin{equation}}
\def\ee{\end{equation}}
\def\bea{\begin{eqnarray}}
\def\eea{\end{eqnarray}}

\def\g{\gamma} 

\def\e{\epsilon}

\def\bg{\bar{g}}
\def\beq{\begin{eqnarray}}\def\eeq{\end{eqnarray}}
\def\ba#1\ea{\begin{align}#1\end{align}}
\def\bg#1\eg{\begin{gather}#1\end{gather}}
\def\bm#1\em{\begin{multline}#1\end{multline}}
\def\bmd#1\emd{\begin{multlined}#1\end{multlined}}

\def\D{\Delta}
\def\e{\epsilon}

\def\g{\gamma}

\def\pa{\partial}

\def\({\left(}
\def\){\right)}
\def\[{\left[}
\def\]{\right]}

\def\g{\gamma}

\def\D{\Delta}

\pdfoutput=1

\title{Conformal properties of soft-operators - 2 : Use of null-states}
\author{Shamik Banerjee, Pranjal Pandey}
\affiliation{Institute of Physics, \\ Sachivalaya Marg, Bhubaneshwar, India-751005 \\ and \\ Homi Bhabha National Institute, Anushakti Nagar, Mumbai, India-400085}

\emailAdd{banerjeeshamik.phy@gmail.com, pranofmpvm@gmail.com}

\abstract {Representations of the (Lorentz) conformal group with the soft operators as highest weight vectors have two universal properties, which we clearly state in this paper. Given a soft operator with a certain dimension and spin, the first property is about the existence of "(large) gauge transformation" that acts on the soft operator. The second property is the decoupling of (large) gauge-invariant null-states of the soft operators from the $S$-matrix elements. In each case, the decoupling equation has the form of zero field-strength condition with the soft operator as the (gauge) potential. Null-state decoupling effectively reduces the number of polarisation states of the soft particle and is crucial in deriving soft-theorems from the Ward identities of asymptotic symmetries. To the best of our understanding, these properties are not directly related to the Lorentz invariance of the $S$-matrix or the existence of asymptotic symmetries. We also verify that the results obtained from the decoupling of null-states are consistent with the leading and subleading soft-theorems with finite energy massive and massless particles in the external legs.}

\begin{document}
%\preprint{}
\maketitle
%\tableofcontents
\flushbottom
\section{Soft-operators}
Soft operators are central in the study of the relationship between soft theorems and asymptotic symmetries in flat space-time \cite{Strominger:2013lka, Strominger:2013jfa, He, Strominger:2014pwa, He:2014cra, Kapec:2015ena,Kapec:2014zla,Kapec:2015vwa,Pate:2017fgt,Kapec:2016jld,Campiglia:2015qka, Campiglia:2015kxa,Bondi:1962px, Strominger:2017zoo,Banerjee:2019aoy}. These operators live on the celestial sphere $S^{D-2}$ and transform as primaries of the Lorentz group $SO(D-1,1)$, which acts on the celestial sphere by conformal transformation. In this paper, following \cite{Banerjee:2019aoy}, we continue our study of the properties of the conformal representation with a soft operator as the highest weight state. 

For our purpose we consider primary operators $\{O^{\D}_{R}(x)\}$ which transform in the representation $R$ of the rotation group $SO(D-2)$ and has conformal dimension $\D$. Here $x$ denotes the Cartesian coordinates of $R^{D-2}$ which is the stereographic image of the celestial sphere $S^{D-2}$. To maximise our freedom we do not restrict the representation $R$ or the dimension $\D$ in any way.

Now, the origin of a subset of the primary operators $\{O^{\D}_{R}(x)\}$ with integer $\D$ and transforming in the symmetric traceless tensor representation of $SO(D-2)$, can be explained in the following way. Suppose we formally expand the physical (creation) annihilation operator of a massless particle of integer spin-$l$ in a Laurent series as,\footnote{The null momentum of the massless particle is parametrized as, $p^{\mu}(\omega, \vec x) = \omega (1 + \vec x^2 , 2 \vec x, 1-\vec x^2)$. For details please see the appendix.}
\be\label{le}
A_{a_1...a_l}(\omega,x) = \sum_{p\in\mathbb{Z}} \frac{S^p_{a_1...a_l}(x)}{\omega^p}
\ee
Now, using the Lorentz transformation property of $A_{a_1...a_l}(\omega, x)$ it is easy to check that the operator $S^p_{a_1...a_l}(x)$ transforms like a conformal primary operator of scaling dimension $\D=p$ and spin-$l$. $S^p_{a_1...a_l}(x)$ is an example of a spin-$l$ soft operator and a member of the set $\{O^{\D}_{R}(x)\}$. 

Let us now substitute this expansion in an $S$-matrix element and for concreteness let us take $l=1$, 
\be
\bra{out} A_{a}(\omega,x)\ket{in} = \sum_{p\in\mathbb{Z}} \frac{\bra{out}S^p_{a}(x)\ket{in}}{\omega^p}
\ee
Let us emphasise that \emph{the $\bra{out}$ and the $\ket{in}$ states may contain both massive and massless particles}. Now from the calculation of $S$-matrix elements we know that : 

1) $\bra{out}S^p_{a}(x)\ket{in} = 0$ for $p>1$. In other words, the primary operators $S^{p>1}_{a_1...a_l}(x)$ decouple from the $S$-matrix elements. 

2) The $(\D=1,l=1)$ primary operator $S^1_a(x)$ is the leading soft-photon operator. 

3) Similarly the $(\D=0,l=1)$ primary operator $S^0_a(x)$ is the subleading soft-photon operator. 

4) The \emph{primary} operators $S^{p}_a(x)$ with $\D=p<0$ may also be called soft-operators, although, this does \emph{not} (necessarily) imply factorisation of the matrix element $\bra{out}S^p_{a}(x)\ket{in}$.  

The same discussion applies to particles of higher spin. 

So we can see that some of the primary operators have interpretation as soft-operators arising from \emph{physical} creation-annihilation operators of massless gauge particles, but, for our purpose, we have to assume the existence of more general primary operators which do not have any such interpretation. We will see that they play the role of "transformation parameters".  

Before we leave this section we would like to mention that in $D=4$ the (formal) expansion \eqref{le} requires some intersting modification due to the existence of the logarithmic terms \cite{Sahoo:2018lxl,Campiglia:2019wxe} in the soft expansion. But we leave this for future work. 

%Moreover, the are some evidences which show that there may exist an Euclidean CFT on the celestial sphere which holographically computes the $S$-matrix in asymptotically flat space-time \cite{Kapec:2016jld,Kapec:2017gsg,Pasterski:2016qvg,Pasterski:2017kqt,Pasterski:2017ylz,Cheung:2016iub,deBoer:2003vf,Banerjee:2018gce,Banerjee:2018fgd,Cardona:2017keg,Lam:2017ofc,Banerjee:2017jeg,Schreiber:2017jsr,Donnay:2018neh,Pate:2019mfs,Adamo:2019ipt,Puhm:2019zbl,Bagchi:2016bcd}.

%This is a formal expansion which is expected to be valid in space-time dimensions $D > 4$. In $D=4$ there can be logarithmic terms in the expansion. 

\section{Universal properties of representation of the conformal group with soft-operator as the highest weight vector}

In \cite{Banerjee:2019aoy} it was shown that Lorentz invariance of the $S$-matrix, together with the Ward-identities of the asymptotic symmetries, require the soft-operators to satisfy certain differential equations. For example, the leading soft-photon operator $S_a(x) (= S^1_a(x))$ was shown to satisfy the (Euclidean) Maxwell's equation, $\pa_a(\pa_aS_b - \pa_bS_a)=0$. Similarly, the leading soft-graviton operator $h_{ab}(x) (=S^1_{ab}(x))$, in $D=6$, satisfies the equation,
\be\label{ee}
\pa^2 h_{ab} - \frac{2}{3} \bigg( \pa_a \pa_c h_{cb} + \pa_b \pa_c h_{ca} \bigg) +\frac{1}{3} \delta_{ab} \pa_c\pa_d h_{cd} = 0
\ee 
These equations are additional constraints which need to be satisfied if we want to derive the soft-theorems from Ward-Identities of the asymptotic symmetries. 

Now, the curious fact is that both these equations are equations of (Euclidean) gauge theories. Maxwell's equations have obvious gauge invariance whereas the second equation \eqref{ee} is invariant under the transformation \cite{Erdmenger:1997wy},
\be
h_{ab}(x) \rightarrow h_{ab}(x) + \bigg( \pa_a\pa_b - \frac{1}{4} \delta_{ab} \pa^2 \bigg) \phi(x)
\ee
where $\phi(x)$ is a $(\D=-1,l=0)$ primary. 

In this paper, \emph{instead of Lorentz invariance and Ward-identity, we take this "gauge-structure" as the starting point.} It turns out that the conformal representation built on any soft-operator $S^{\D}_{a_1....a_l}(x)$ obeys two simple \emph{universal} rules stated below,

%We formulate two heuristic rules using the familiar language of conformal representation theory. Simply stated, the consequence of these rules is to replace the "equation of motion of the gauge-theory" with the (stronger) condition of vanishing "field strength". This condition is more natural from the point of view of the conformal representation theory and technically much simpler. Let us now state the rules :  

1) \underline{\bf{Symmetry transformation}} :  

Given a $(\D,l)$ primary field $S^{\D}_{a_1....a_l}(x)$, we look for another primary field $O^{\delta}_{R}(x)$ which transforms in the representation $R$ of the rotation group $SO(D-2)$ and has dimension $\delta<\D$. The \emph{defining property of the primary $O^{\delta}_{R}(x)$ is that it should have a primary descendant of weight $\D$ and spin $l$}. The primary field $O^{\delta}_{R}(x)$ may or may not exist depending on the quantum numbers $(\D,l)$. Let us consider the interesting case where it exists and denote the corresponding primary descendant by $N^{\D}_{a_1....a_l}( x)$.  

The existence of the $(\D,l)$ primary descendant $N^{\D}_{a_1....a_l}( x)$ allows us to define a "global symmetry transformation" by,
\be\label{gs}
\boxed{
S^{\D}_{a_1....a_l}( x) \rightarrow S^{\D}_{a_1....a_l}(x) + N^{\D}_{a_1....a_l}(x)}
\ee

Since $N^{\D}_{a_1....a_l}(x)$ can be written as a differential operator acting on the primary $O^{\delta}_{R}(x)$, one can think of $O^{\delta}_{R}(x)$ as the transformation function or parameter. 

2) \underline{\bf{Decoupling of Null-states}} : 
The next step is to construct the conformal representation with the soft operator $S^{\D}_{a_1....a_l}(x)$ as the highest-weight vector. For our purpose, the interesting states in the representation space are the primary descendants. The primary descendants have two possible fates, which we now describe:  

\underline{\bf{Case-A}} $(\D,l)$ is such that the primary $O^{\delta}_{R}(x)$ does \emph{not} exist. In that case, \emph{we cannot set any of the primary descendants of $S^{\D}_{a_1....a_l}(x)$ to zero.} 

\underline{\bf{Case-B}} $(\D,l)$ is such that the primary field $O^{\delta}_{R}(x)$ exists and the symmetry transformation \eqref{gs} is defined. In this case, we can \emph{set \underline{only} those primary descendants of $S^{\D}_{a_1....a_l}(x)$ to zero which are \underline{invariant} under the symmetry transformation \eqref{gs}.} In other words, the primary descendants of $S^{\D}_{a_1....a_l}(x)$, which are invariant under the symmetry transformation \eqref{gs} \emph{decouple from the $S$-matrix element}. 

Lorentz invariance of the $S$-matrix has no obvious connection to the decoupling. Without decoupling there will be a mismatch between the number of Ward-identities of the asymptotic symmetries and the number of soft $S$-matrix elements. So, in order to derive soft-theorems from the corresponding Ward-identities, decoupling is crucial. It seems that the correct explanation of the \emph{universal} rules we have stated, goes beyond the "asymptotic symmetry - soft theorem" correspondence. In fact there are some indications, that an Euclidean CFT living on the celestial sphere holographically computes the $S$-matrix elements in the asymptotically flat space \cite{Kapec:2016jld,Kapec:2017gsg,Pasterski:2016qvg,Pasterski:2017kqt,Pasterski:2017ylz,Cheung:2016iub,deBoer:2003vf,Banerjee:2018gce,Banerjee:2018fgd,Cardona:2017keg,Lam:2017ofc,Banerjee:2017jeg,Schreiber:2017jsr,Donnay:2018neh,Fan:2019emx,Pate:2019mfs,Adamo:2019ipt,Puhm:2019zbl,Ball:2019atb, Bagchi:2016bcd}. If this is true then the dual CFT may provide a \textit{microscopic} explanation of the universality. Another avenue will be to understand the asymptotic symmetries in flat space-time from the string theory point of view along the line of \cite{Giveon:1998ns,Geyer:2014lca,Lipstein:2015rxa,Ball:2019atb,Adamo:2019ipt}. We leave these questions for future study.   

In the rest of the paper we derive consequences of this decoupling and check their consistency with soft-theorems. This demonstration also shows that the above mentioned properties of the gauge particles are intrinsic and do not depend on whether the rest of the particles in the scattering process are massive or massless.  

\section{Consequences of Decoupling}

In this section we study the consequences of decoupling for various soft operators. We will assume throughout the rest of the paper that the space-time dimension $D=n+2$ is even. 

\subsection{Leading soft-photon or infinite-dimensional $U(1)$} 
 
The leading soft-photon $S_a(x)$ is a $(\D=1,l=1)$ conformal primary and it has a primary descendant at level-$1$, given by, 
\be
F_{ab} = \pa_a S_b - \pa_b S_a
\ee 

The symmetry transformation acting on $S_a$ is given by, \footnote{See the appendix for a discussion of this.}
\be
S_a \rightarrow S_a + \pa_a\phi
\ee
where the transformation parameter $\phi$ is a $(\D=0,l=1)$ conformal primary. The null state $F_{ab}$ is invariant under this transformation and so we set it to zero, i.e, 
\be
\boxed{
F_{ab} = \pa_a S_b - \pa_b S_a =0} 
\ee

If we think of this as a classical field equation then this has solution,
\be
S_a(x) = \pa_a \chi(x)
\ee

Now, at the level of $S$-matrix this implies that,
\be\label{pg1}
\bra{\{p^{out}\}} S_a(x) \ket{\{p^{in}\}} = \pa_a \chi(x,\{p^{out}\},\{p^{in}\})
\ee
So the $S$-matrix elements corresponding to different polarisations of the soft-photon are determined in terms of a \emph{single scalar function} $\chi$. So, the \emph{decoupling of null-states $F_{ab}(x)$ effectively reduces the number of polarization states of the soft-photon from $n=D-2$ to $1$}. This is consistent with the fact that the transformation parameter $\phi(x)$ is a scalar. 

\subsection{Leading soft-graviton or Supertranslation}

\subsubsection{$n=D-2>2$}

The leading soft-graviton is a $(\D=1,l=2)$ conformal primary which we denote by $S_{ab}(x)$. For $D>4$, $S_{ab}(x)$ has a unique level-$1$ null-state given by,
\be
\boxed{
O_{abc}(x) = \pa_a S_{bc} - \pa_b S_{ac} + \frac{1}{n-1} \big( \delta_{bc} \pa_d S_{da} - \delta_{ac} \pa_d S_{db} \big)}
\ee 
The operator $O_{abc}$ satisfies,
\be
O_{abc} = - O_{bac} , \quad O_{aba} = O_{abb} = 0
\ee
The symmetry transformation acting on $S_{ab}$ is given by,
\be
\boxed{
S_{ab} \rightarrow S_{ab} +  \big( \pa_a\pa_b - \frac{1}{n} \delta_{ab} \pa^2 \big) f(x)}
\ee
where the transformation parameter $f(x)$ is a $(\D=-1,l=0)$ conformal primary. The null-state $O_{abc}(x)$ is invariant under this transformation, i.e, 
\be
O\big[ S_{ab} \big] = O \big[ S_{ab} + \big( \pa_a\pa_b - \frac{1}{n} \delta_{ab} \pa^2 \big) f(x) \big]
\ee
So we have to set the null-state $O_{abc}$ to zero, i.e, 
\be\label{EE}
\boxed{
O_{abc} (x) = \pa_a S_{bc} - \pa_b S_{ac} + \frac{1}{n-1} \big( \delta_{bc} \pa_d S_{da} - \delta_{ac} \pa_d S_{db} \big) = 0}
\ee

Thinking of this as a classical field equation, let us now show that,
\be
\boxed{
O_{abc} (x) = 0 \Rightarrow S_{ab}(x) = \big( \pa_a\pa_b - \frac{1}{n} \delta_{ab} \pa^2 \big) F(x)}
\ee
for some function $F(x)$. We now give a formal proof of this. \\

\underline{\bf{Proof}} :
 We have to solve the equation
\be\label{meq}
O_{abc}(x) = \pa_a S_{bc} - \pa_b S_{ac} + \frac{1}{n-1} \big( \delta_{bc} \pa_d S_{da} - \delta_{ac} \pa_d S_{db} \big) = 0
\ee
Now,
\be\label{max}
\pa_c O_{abc}(x) = 0 \Rightarrow \pa_a\pa_c S_{cb} - \pa_b\pa_c S_{ca} = 0
\ee
Define
\be
V_a = \pa_b S_{ab}
\ee
In terms of $V_a$, \eqref{max} becomes
\be\label{flat}
\pa_a V_b - \pa_b V_a = 0 \Rightarrow \boxed{V_a = \pa_b S_{ab} = \pa_a \phi(x)}
\ee

Now we also have, 
\be\label{1}
\pa_a O_{abc}(x) = \pa^2 S_{bc} - \frac{n}{n-1} \big(\pa_b\pa_c - \frac{1}{n} \delta_{bc} \pa^2 \big) \phi(x) = 0 
\ee
where we have used \eqref{flat}. The solution of \eqref{1} can be written as,
\be\label{lap}
\boxed{
S_{ab} =  \frac{n}{n-1} \big(\pa_a\pa_b - \frac{1}{n} \delta_{ab} \pa^2 \big) \psi(x) + \tilde S_{ab}(x)} 
\ee
where,
\be
\pa^2 \psi = \phi , \quad \pa^2\tilde S_{ab}(x) = 0 , \quad \pa_a\tilde S_{ab}(x) = 0 
\ee
Now substituting \eqref{lap} into \eqref{meq} we get,
\be\label{Flat}
\boxed{
\pa_a \tilde S_{bc} - \pa_b \tilde S_{ac} = 0}
\ee
where we have used $\pa_a \tilde S_{ab}=0$. The general solution of \eqref{Flat} can be written as,
\be
\tilde S_{ab} = \pa_a \xi_b(x)
\ee
where $\xi_a$ is some vector field. Now we impose the constraints that $\tilde S_{ab}$ is symmetric and traceless. This gives us,
\be
\tilde S_{ab} = \tilde S_{ba} \Rightarrow \pa_a \xi_b - \pa_b \xi_a = 0 \Rightarrow \xi_a(x) = \pa_a \chi(x) 
\ee 
for some scalar field $\chi(x)$. Therefore,
\be
\tilde S_{ab}(x) = \pa_a\pa_b\chi(x) 
\ee
Tracelessness of $\tilde S_{ab}(x)$ further implies that 
\be
\pa^2 \chi(x) = 0
\ee
Therefore we can write,
\be
\tilde S_{ab}(x) = \big( \pa_a \pa_b - \frac{1}{n} \delta_{ab}\pa^2\big) \chi(x) 
\ee

Now if we define
\be
F(x) = \frac{n}{n-1} \psi(x) + \chi(x)
\ee
then we can write,
\be
\boxed{
S_{ab}(x) = \big( \pa_a \pa_b - \frac{1}{n} \delta_{ab}\pa^2\big) F(x)}  
\ee
So $S_{ab}(x)$ has the "pure-gauge" form and $O_{abc}(x)$ can be thought of as the "field-strength" for the Supertranslation transformation.

Now at the level of $S$-matrix this implies that,
\be\label{pg2}
\boxed{
\bra{\{p^{out}\}} S_{ab}(x) \ket{\{p^{in}\}} = \big( \pa_a \pa_b - \frac{1}{n} \delta_{ab}\pa^2\big) \psi(x,\{p^{out}\},\{p^{in}\})}
\ee
So again we can see that the decoupling of the null-state effectively reduces the number of polarization states of soft-graviton to $1$.

\subsubsection{$D=4$}

In this case there are two leading soft gravitons with scaling dimension and spin given by $(\D=1,l=\pm 2)$. They are $SL(2,\mathbb{C})$ primaries with weights given by, $(h=3/2, \bar h = -1/2)$ and $(h=-1/2,\bar h=3/2)$, respectively. Let us denote the primaries by $S_{zz}(z,\bar z)$ and $S_{\bar z\bar z}(z,\bar z)$, respectively. Now one can check that there are two primary descendants or null-states at level-$2$ given by, 
\be
\boxed{
O_1(z,\bar z) = \pa^2_{\bar z} S_{zz} , \quad  O_2(z,\bar z) = \pa^2_{z} S_{\bar z\bar z}}
\ee

Now the symmetry transformation acting on the soft-gravitons is given by,
\be
\boxed{
S_{zz} \rightarrow S_{zz} + \pa^2_z f(z,\bar z), \quad S_{\bar z\bar z} \rightarrow S_{\bar z\bar z} + \pa^2_{\bar z} f(z,\bar z)}
\ee
where $f(z,\bar z)$ is a $SL(2,\mathbb{C})$ primary of weight $(h=-1/2, \bar h = -1/2)$. Now the unique linear combinations of the null-states $O_1$ and $O_2$, which is invariant under the symmetry transformation, is given by,
\be
O(z,\bar z) = O_1(z,\bar z) - O_2(z,\bar z) = \pa^2_{\bar z} S_{zz} - \pa^2_{z} S_{\bar z\bar z}
\ee
This is the null-state which decouples, i.e,
\be
\boxed{
O(z,\bar z) = \pa^2_{\bar z} S_{zz} - \pa^2_{z} S_{\bar z\bar z} =0} 
\ee

\subsection{Subleading soft graviton or Superrotation}

\subsubsection{$D=4,n=2$}
In $D=4$ there are two subleading soft gravitons with scaling dimension $\D=0$ and spin $l=\pm 2$. They are $SL(2,\mathbb{C})$ primaries with weights given by, $(h=1,\bar h=-1)$ and $(h=-1, \bar h=1)$, respectively. We denote them by $S_{zz}$ and $S_{\bar z\bar z}$, respectively. Now there are two null-states at level-$3$, given by,
\be
\boxed{
O^{(1)}_{\bar z}(z,\bar z) = \pa^3_{\bar z} S_{zz}, \quad O^{(2)}_{z}(z,\bar z) = \pa^3_{z} S_{\bar z\bar z}}
\ee

Now the symmetry transformation acting on the subleading soft-gravitons is given by,
\be\label{ssld}
\boxed{
S_{zz} \rightarrow S_{zz} + 2 \pa_z V_z , \quad S_{\bar z\bar z} \rightarrow S_{\bar z\bar z} + 2 \pa_{\bar z} V_{\bar z}}
\ee
where $V_z$ is a $SL(2,\mathbb{C})$ primary with weight $(h=0,\bar h=-1)$ and $V_{\bar z}$ is the complex conjugate with weight $(h=-1, \bar h=0)$. 

Now we can see that both the primary descendants, $O^{(1)}_{\bar z}$ and $O^{(2)}_z$, are not invariant under the symmetry transformation and also, we cannot take any linear combination of them because they carry different quantum numbers. Therefore, \emph{in $D=4$, no primary descendant of the subleading soft graviton decouples from the $S$-matrix}. 

This is an example where there are null-states but still \emph{they do not decouple from the $S$-matrix elements because none of them are invariant under the symmetry transformation \eqref{ssld}. In other words, we cannot set them equal to zero.} This is consistent with the fact that the transformation parameter is a vector field with two components and graviton also has two polarization states. 

\subsubsection{$D=6,n=4$}

The situation is different in $D=6$ (and higher) dimensions. The subleading soft graviton, denoted by $S_{ab}(x)$ is a conformal primary with $\D=0$ and $l=2$ and the symmetry transformation acting on $S_{ab}$ is given by,
\be
\boxed{
S_{ab} \rightarrow S_{ab} + \pa_a V_b + \pa_b V_a - \frac{1}{2} \delta_{ab} \pa\cdot V}
\ee
where $V_a$ is a $(\D=-1,l=1)$ conformal primary. 

Now the transformation parameter $V_a(x)$ is a vector field with $4$ components and the number of polarization states of the graviton is $9$. So there is a mismatch between the number of Ward-identities and the number of soft $S$-matrix elements. This will be cured if some null-state of $S_{ab}(x)$ decouples from the $S$-matrix elements.  

Now, $S_{ab}(x)$ has a null state at level-$2$ given by,
\bea
W_{abcd} = \pa_{a} \pa_{c} S_{b d}-\pa_{a} \pa_{d} S_{b c}-\pa_{b} \pa_{c} S_{a d}+\pa_{b} \pa_{d} S_{a c}\nonumber \\ 
- \frac{1}{2}\delta_{a c} \pa^2 S_{b d}+\frac{1}{2}\delta_{a d} \pa^2 S_{b c}+\frac{1}{2}\delta_{b c} \pa^2 S_{a d} - \frac{1}{2}\delta_{b d} \pa^2 S_{a c}\nonumber \\
+\frac{1}{2}\delta_{a c} \pa_{b} \pa_{e} S_{d e}+\frac{1}{2}\delta_{a c} \pa_{d} \pa_{e} S_{b e} - \frac{1}{2}\delta_{a d} \pa_{b} \pa_{e} S_{c e} - \frac{1}{2}\delta_{a d} \pa_{c} \pa_{e} S_{b e}\nonumber \\
 - \frac{1}{2}\delta_{b c} \pa_{a} \pa_{e} S_{d e} - \frac{1}{2}\delta_{b c} \pa_{d} \pa_{e} S_{a e}+\frac{1}{2}\delta_{b d} \pa_{a} \pa_{e} S_{c e}+\frac{1}{2}\delta_{b d} \pa_{c} \pa_{e} S_{a e}\nonumber\\
 - \frac{1}{3}\delta_{a c} \delta_{b d} \pa_{e} \pa_{f} S_{e f}+\frac{1}{3}\delta_{a d} \delta_{b c} \pa_{e} \pa_{f} S_{e f}
\eea

As the notation suggests, the null (field) state $W_{abcd}$ is essentially the \emph{linearization of the Weyl tensor} around flat space $\delta_{ab}$ and $S_{ab}$ is the trace-free part of the "metric fluctuation". 

Now it is obvious that,
\be
W\big[S_{ab}\big] = W \big[S_{ab} + \pa_a V_b + \pa_b V_a - \frac{1}{2} \delta_{ab} \pa\cdot V\big]
\ee
Therefore we have to set the null state $W_{abcd}$ to zero, i.e, 
\be\label{weylde}
\boxed{
W_{abcd} \big[S\big] = 0}
\ee
is the decoupling equation. Thought of as a classical field equation this has the solution,
\be
\boxed{
S_{ab} = \pa_a \xi_b + \pa_b \xi_a - \frac{1}{2} \delta_{ab} \pa\cdot \xi }
\ee
Here we have taken into account that $S_{ab}$ is traceless. 

At the level of $S$-matrix this implies that,
\be\label{pg3}
\boxed{
\bra{\{p^{out}\}} S_{ab}(x) \ket{\{p^{in}\}} = \pa_a \psi_b(x,\{p^{out}\},\{p^{in}\}) + \pa_b \psi_a(x,\{p^{out}\},\{p^{in}\}) - \frac{1}{2} \delta_{ab} \pa\cdot \psi(x,\{p^{out}\},\{p^{in}\})}
\ee
So we can see that in $D=6$, the decoupling of the null state $W_{abcd}$, effectively reduces the $9$ polarisation states of the subleading soft-graviton to $4$. 

\section{Consistency with soft-theorems : Massive case}
In this section we will explicitly verify that the forms of the soft $S$-matrix elements obtained from the decoupling equations match with that obtained from the soft theorems. We work with massive external states. Massless case can be studied by taking appropriate limit \cite{Pasterski:2017kqt}. 

In this section we have defined $\eta_k = +1$ for the $k$-th outgoing particle and $\eta_k=-1$ for the $k$-th incoming particle. We have also taken the soft particle to be outgoing and for simplicity we have assumed the external states to be scalars. The parametrization of the momentum of a massive particle is discussed in the appendix. 

\subsection{Leading soft photon or U(1)} 
In terms of the leading soft-photon operator $S_a(\vec w)$, Weinberg's soft photon theorem \cite{Weinberg:1964ew,Weinberg:1965nx} can be written as,
\be
\begin{gathered}
\bra{\{y_i,\vec x_i,Q_i\},out} S_a(\vec w) \ket{\{y_j,\vec x_j,Q_j\},in} \nonumber  \\ 
= \bigg[ \sum_k \eta_k Q_k \frac{p(y_k,\vec x_k) \cdot \epsilon_a(\vec w)}{p(y_k,\vec x_k)\cdot \hat q(\vec w)} \bigg] \bra{\{y_i,\vec x_i,Q_i\},out}\ket{\{y_j,\vec x_j,Q_j\},in}
\end{gathered}
\ee
Here $Q_k$ is the charge of the $k$-th particle and $\epsilon_a(\vec w)$ is the polarization vector of the soft photon. 

The soft-factor can be calculated to be, 
\be
\begin{gathered}
 %\sum_{i\in out} Q_i \frac{w_a - x_{ia}}{y_i^2 + |\vec w - \vec x_i|^2} - \sum_{j\in in} Q_j \frac{w_a - x_{ja}}{y_j^2 + |\vec w - \vec x_j|^2} \\ 
\bigg[ \sum_k \eta_k Q_k \frac{p(y_k,\vec x_k) \cdot \epsilon_a(\vec w)}{p(y_k,\vec x_k)\cdot \hat q(\vec w)} \bigg] = \frac{\pa}{\pa w_a} \bigg[ \sum_k\eta_k \frac{1}{2} Q_k \ln(y_k^2 + |\vec w - \vec x_k|^2)\bigg] \\
%=  \frac{\pa}{\pa w_a} \Lambda(\vec w , \{y, \vec x, Q\})
\end{gathered}
\ee
Therefore the matrix element of the soft-photon operator can be written as, 
\be\label{td}
\boxed{
\bra{\{y_i,\vec x_i,Q_i\},out} S_a(\vec w) \ket{\{y_j,\vec x_j,Q_j\},in} = \frac{\pa}{\pa w_a} \Lambda(\vec w , \{y, \vec x, Q\})}
\ee
where,
\be
\begin{gathered}
\Lambda(\vec w , \{y, \vec x, Q\}) = \bigg[ \sum_k \eta_k\frac{1}{2} Q_k \ln(y_k^2 + |\vec w - \vec x_k|^2)\bigg] \bra{\{y_i,\vec x_i,Q_i\},out}\ket{\{y_j,\vec x_j,Q_j\},in}
\end{gathered}
\ee

\eqref{td} matches with the form \eqref{pg1} of the leading soft photon theorem obtained from the decoupling of the null-state. 

\subsection{Leading soft graviton or Supertranslation}
In terms of the leading soft graviton operator $S_{ab}(\vec w)$, Weinberg's soft graviton theorem \cite{Weinberg:1964ew,Weinberg:1965nx} can be written as, 
\be
\begin{gathered}
\bra{\{y_i,\vec x_i\},out} S_{ab}(\vec w) \ket{\{y_j,\vec x_j\},in} \nonumber  = \bigg[ \sum_k \eta_k  \frac{p_\mu(y_k,\vec x_k) p_\nu(y_k,\vec x_k) \epsilon^{\mu\nu}_{ab}(\vec w)}{p(y_k,\vec x_k)\cdot \hat q(\vec w)}\bigg] \bra{\{y_i,\vec x_i\},out}\ket{\{y_j,\vec x_j\},in}
\end{gathered}
\ee
where $\epsilon^{\mu\nu}_{ab}(\vec w)$ is the graviton polarization vector. 

Now the soft-factor can be computed to be, 
\be
\begin{gathered}
\gamma \bigg[ \sum_k \eta_k \frac{m_k}{y_k} \bigg(\delta_{ab} - n \frac{(w_a-x_{ka})(w_b-x_{kb})}{y_k^2 + |\vec w - \vec x_k|^2}\bigg) \bigg]  = \bigg(\pa_a^w\pa_b^w - \frac{1}{n} \delta_{ab} \pa_w^2 \bigg) F(\vec w , \{y, \vec x, m\})
\end{gathered}
\ee
where,
\be
\boxed{
\begin{gathered}
F(\vec w , \{y, \vec x, m\}) = 
\frac{-n\gamma}{2} \bigg[ \sum_k \eta_k m_k \frac{y_k^2 + |\vec w - \vec x_k|^2}{2y_k} \ln(y_k^2 + |\vec w - \vec x_k|^2)\bigg]
\end{gathered}}
\ee
Here $m_k$ is the mass of the $k$-th particle and $\gamma$ is a numerical constant whose precise value is not important for us. Now, using this we can write,
\be\label{gth}
\boxed{
\bra{\{y_i,\vec x_i\},out} S_{ab}(\vec w) \ket{\{y_j,\vec x_j\},in} = \bigg(\pa_a^w\pa_b^w - \frac{1}{n} \delta_{ab} \pa_w^2 \bigg) \Lambda(\vec w , \{y, \vec x, m\})}
\ee
where we have defined,
\be
\Lambda(\vec w , \{y, \vec x, m\}) = F(\vec w , \{y, \vec x, m\}) \bra{\{y_i,\vec x_i\},out}\ket{\{y_j,\vec x_j\},in}
\ee

\eqref{gth} matches with the form \eqref{pg2} of the leading soft graviton theorem obtained from the decoupling of the null-state

\subsection{Subleading soft graviton theorem in $D=6 (n=4)$ or Superrotation}
In terms of the subleading soft graviton operator - again denoted by $S_{ab}(\vec w)$ - Cachazo-Strominger subleading soft graviton theorem \cite{Cachazo:2014fwa} can be written as, 
\be
\begin{gathered}
\bra{\{y_i,\vec x_i\},out} S_{ab}(\vec w) \ket{\{y_j,\vec x_j\},in} \\ = -i \sum_k \eta_k \frac{\epsilon_{ab}^{\mu\nu}(\vec w)p^k_{\mu}q^{\rho}}{p^k\cdot q} J^k_{\rho\nu} \bra{\{y_i,\vec x_i\},out}\ket{\{y_j,\vec x_j\},in}
\end{gathered}
\ee
Since all the external particles are scalars, the (orbital) angular momentum operator is given by,
\be
J^k_{\rho \nu}=-i\(p^k_{\rho}\frac{\pa}{\pa p_k^\nu}-p^k_{\nu}\frac{\pa}{\pa p_k^\rho}\)
\ee

Now a straightforward calculation gives, 
\be\label{weyl}
\boxed{
\bra{\{y_i,\vec x_i\},out} S_{ab}(\vec w) \ket{\{y_j,\vec x_j\},in} = \pa^w _a\zeta_b+\pa^w_b \zeta_a-\frac{1}{2}\delta_{ab}\pa^w .\zeta}
\ee
where the vector $\zeta$ is given by,
\be
\begin{gathered}
\zeta_a(\vec{w},\{y,\vec{x}\})=\frac{\g}{2}\sum_k \eta_k\bigg[ \ln(y_k^2+|\vec{w}-\vec{x_k}|^2)\bigg \{-(w_a-x_{ka})y_k\frac{\pa}{\pa y_k} \\ +\left((w_a-x_{ka})(w_c-x_{kc})-\frac{1}{2}(y_k^2+|\vec{w}-\vec{x_k}|^2)\delta_{ac}\right)\frac{\pa}{\pa x_{kc}}\bigg\}\bra{\{y_i,\vec x_i\},out}\ket{\{y_j,\vec x_j\},in} \bigg]
\end{gathered}
\ee
Here $\gamma$ is a numerical constant whose exact value is not important for us. 

\eqref{weyl} matches with the form \eqref{pg3} of the subleading soft-graviton theorem obtained from the decoupling of the null-state.

\section{Comments on the construction of the soft-charge}
In \emph{even} space-time dimensions the \emph{soft-charge} \cite{Strominger:2013lka, Strominger:2013jfa, He, Strominger:2014pwa, He:2014cra, Kapec:2015ena,Kapec:2014zla,Kapec:2015vwa,Pate:2017fgt,Kapec:2016jld,Campiglia:2015qka, Campiglia:2015kxa,Bondi:1962px, Strominger:2017zoo} corresponding to a particular symmetry transformation can always be written as,
\be\label{sc}
Q_S[\xi] = \int d^{D-2}{x} \ \xi(x)\cdot O(x)
\ee
where $\xi(x)$ is the transformation parameter and $O(x)$ is a local operator, which is also a \emph{primary} of the (Lorentz) conformal group \cite{Banerjee:2019aoy}. The tensor indices of $O(x)$ are dual to those of $\xi(x)$, so that the \emph{contraction $\xi(x)\cdot O(x)$ is a scalar of dimension $(D-2)$}. In even space-time dimension $O(x)$ can be constructed as a descendant of the soft operator. Let us now illustrate this with an example. 

Consider the case of subleading soft-graviton in $D=6$. In this case we know that the transformation parameter is a $(\D=-1,l=1)$ primary which we denote by $\xi_a(x)$. So the soft-charge can be written as, 
\be
Q_S[\xi] = \int d^{4}{x} \ \xi_a(x)O_a(x)
\ee
Since $\xi_a(x)O_a(x)$ should have dimension $4$ and $\xi(x)$ has dimension $\D=-1$, \emph{$O_a(x)$ should be a vector primary with dimension $\D=5$}. Now $O_a(x)$ can be constructed as a descendant of the subleading soft-graviton $S_{ab}(x)$ and is given by,
\be\label{scl}
\boxed{
O_a(x) \sim \pa^2 \pa_a \pa_b \pa_c S_{bc} - \frac{3}{4} \pa^4 \pa_b S_{ab}(x)}
\ee 
Here we have used $\sim$ because the overall normalization of the charge cannot be determined in this way. Now since the subleading soft graviton is a $(\D=0,l=2)$ primary, $O_a(x)$ constructed in this way clearly has dimension $\D=5$ and spin $l=1$. One can check by straightforward calculation that \emph{$O_a(x)$, given in \eqref{scl}, is a primary operator as long as the decoupling equation \eqref{weylde} for the subleading soft graviton $S_{ab}(x)$ holds.}

\section{Comparison with \cite{Banerjee:2019aoy}}
Before we conclude we would like to compare the results of \cite{Banerjee:2019aoy} with the current paper. For the sake of concreteness let us compare the decoupling equations for the leading soft-graviton $S_{ab}(x)$ which is a $(\D=1,l=2)$ primary. 

In our current approach the decoupling equation \eqref{EE} is given by,
\be\label{EE1}
O_{abc} (x) = \pa_a S_{bc} - \pa_b S_{ac} + \frac{1}{3} \big( \delta_{bc} \pa_d S_{da} - \delta_{ac} \pa_d S_{db} \big) = 0
\ee 
where we have specialized to the case of $D=6$ or $n=4$. 

Now in \cite{Banerjee:2019aoy}, starting from the Lorentz invariance of the $S$-matrix and the Ward identity for the supertranslation, we arrived at a somewhat different decoupling equation,
\be\label{ee1}
\tilde O_{ab}= \pa^2 S_{ab} - \frac{2}{3} \bigg( \pa_a \pa_c S_{cb} + \pa_b \pa_c S_{ca} \bigg) +\frac{1}{3} \delta_{ab} \pa_c\pa_d S_{cd} = 0
\ee
Now one can easily check that, 
\be
\boxed{
\tilde O_{bc} = \frac{1}{2} (\pa_a O_{abc} + \pa_a O_{acb})}
\ee
Therefore \emph{$\tilde O_{ab}$ is a descendant of $O_{abc}$} -- in fact a primary descendant -- and the decoupling equation 
\be
O_{abc}=0 \Rightarrow \tilde O_{ab}=0 
\ee
But,
\be
\tilde O_{ab}=0 \nRightarrow \tilde O_{abc}=0
\ee
Therefore \eqref{EE1} is a stronger condition than \eqref{ee1}.   

Now, in the case of leading soft graviton or supertranslation, the operator $O(x)$ appearing in the soft charge must be a $(\D=5,l=0)$ primary in $D=6$. Therefore a potential candidate for $O(x)$ is the unique scalar operator $\pa^2\pa_a\pa_b S_{ab}$ of dimension $\D=5$. In \cite{Banerjee:2019aoy}, the decoupling equation \eqref{ee1} was obtained by demanding that the operator $\pa^2\pa_a\pa_b S_{ab}$ be a primary. Now the same thing can also be achieved by using the decoupling equation \eqref{EE1}. To be more precise, under an infinitesimal special conformal transformation, $O(x)=\pa^2\pa_a\pa_b S_{ab}$ transforms as \cite{Banerjee:2019aoy},
\be\label{sct}
O'(x') = \left( 1 + 10 \, \e \cdot x \right) O(x) + \boxed{8 \, \e_a \pa_c\left( \pa_a \pa_b S_{bc}(x) - \pa_b \pa_c S_{ab}(x)\right)}
\ee 
Now,
\be
O_{abc} =0 \Rightarrow \pa_c O_{abc}(x) = 0 \Rightarrow \pa_a\pa_c S_{cb} - \pa_b\pa_c S_{ca} = 0
\ee
Therefore, owing to the decoupling equation \eqref{EE1}, the inhomogeneous term in \eqref{sct} vanish and $\pa^2\pa_a\pa_b S_{ab}$ indeed transforms like a primary. 

%But, as we have shown in \cite{Banerjee:2019aoy}, it is also true that the additional term still vanishes if, instead of \eqref{EE1}, we use the decoupling equation \eqref{ee1}. 

Although the two formulations seem to be mutually consistent, the current one is more general and cannot be simply derived from Lorentz invariance and the existence of infinite dimensional asymptotic symmetries. For example, the decoupling equation \eqref{EE1} of this paper is \emph{sufficient} for the Lorentz invariance of the supertranslation Ward identity. Now, it is natural to conjecture that this is also a \emph{necessary} condition, although the reason may not be Lorentz invariance and asymptotic symmetry. In fact, the current formulation is very natural from a dual CFT perspective, if there is one. We hope to return to these interesting questions in future.  

\section{Acknowledgement}
We would like to thank  all the participants of the $5$-th Indo-Israeli meeting in Nazareth, especially, Abhijit Gadde and Joao Penedones for very helpful discussions. We would also like to thank the participants of the meeting "Stringy days IV : SOFT Holography" in IISER Pune, especially, Alok Laddha and Ashoke Sen for very helpful questions and discussions. Some of the calculation in this paper were done using \cite{Peeters:2007wn}.

\section{Appendix}

\subsection{Notation and conventions}
For the convenience of the reader we collect some useful formulas here. For details we refer to \cite{Kapec:2017gsg,Pasterski:2017kqt}. 

%In this paper we work exclusively with massless particles and \emph{even} space-time dimensions. 

In $D$ dimensional Minkowski space-time we parametrize the null momentum $p^{\mu}(\omega, \vec x)$ of a massless particle as, 
\be
p^{\mu} = \omega (1 + \vec x^2 , 2 \vec x, 1-\vec x^2), \quad  \vec x\in R^{D-2} = R^n
\ee
where $\omega$ is a real number. The Lorentz group $SO(D-1,1)=SO(n+1,1)$ acts on $R^n$ as the group of conformal transformations. The corresponding transformation of $\omega$ is given by, 
\be
\omega' = \bigg| \frac{\pa \vec x'}{\pa \vec x} \bigg| ^{-\frac{1}{n}} \omega
\ee 

We parametrize the energy-momentum vector of a massive particle of mass $m$ as \cite{Pasterski:2016qvg,Pasterski:2017kqt},
\be
p^{\mu} (y,\vec x) = m \bigg( \frac{1+y^2 +\vec x^2}{2y} , \frac{\vec x}{y} , \frac{1 - y^2 - \vec x^2}{2y} \bigg)
\ee
Here $(y,\vec x)$ are the Poincare coordinates on the unit (mass) hyperboloid, $p^2 = -1$. So $\vec x$ are the Cartesian coordinates of the (conformal) boundary $R^{n}$ at infinity, on which the soft-operators live. 

In the rest of the paper we omit the vector sign on $\vec x$ and simply denote it by $x$.

\subsection{Symmetry Transformations}
In this section we will see which primary fields $O_R ^\delta (x)$ can give a null field $N^\Delta_{a_1...a_l}$. The  $(\Delta,l)$ we are interested in are $(1,1),(1,2)$ and $(0,2)$ corresponding to leading photon, leading graviton and subleading graviton respectively. If $O_R ^\delta (x)$ has a primary descendant in the irreducible representation $I$ of $SO(n)$ at level $N$ then the following necessary but not sufficient condition, valid for $n>2$, is satisfied,

\be
\delta=\frac{1}{2N}\left(C_R +N C_V-N(N-1)-C_I\right)\label{condition1}\quad,\quad N\in \{1,2,3...\}
\ee
Here $C_R,C_V (vector)$ and $C_I$ are the quadratic Casimirs for respective representations of $SO(n)$. This formula can be obtained by a simple extension of the method by \cite{osborn}.

If representations $R$ and $I$ are symmetric-traceless then this condition is consistent with the necessary and also sufficient constraints in \cite{Penedones:2015aga}.
Since we are interested in primary descendants $(\Delta,l)$ so 
\be
\delta=\Delta -N~~and~~C_I=l(l+n-2)
\ee
The derivatives generating the descendant are symmetric so $O_R ^\delta (x)$ can give a spin $l=1~or~2$ descendant only for the representations in Figure~\ref{fig11}.
\begin{figure}
	\ytableausetup{smalltableaux}
	\be
	\text{for }l=1:\quad
	\begin{ytableau}
	1 & 2 &	3 & ... & s
	\end{ytableau}~,\quad
	\begin{ytableau}
		1&2&3&..&i\\1
	\end{ytableau}~~,i\geq 1\nonumber
\ee
\be
\text{for } l=2:\quad
\begin{ytableau}
	1 & 2 &	3 & ... & s
\end{ytableau}~,\quad
\begin{ytableau}
	1&2&3&..&i\\1
\end{ytableau}~,\quad
\begin{ytableau}
	1&2&3&..&i\\1&2
\end{ytableau}~~,i\geq 2\nonumber
\ee

\caption{Representations R of $O_R ^\delta (x)$ which can have a descendant of spin-\textit{l}}
\label{fig11}
\end{figure}
Casimir for representation $R_{ij}$ is
\bea
R_{ij}: 	\ytableausetup{smalltableaux}\begin{ytableau}
	1&2&3&.. &..  &..&i\\1&2&..&j
\end{ytableau}\nonumber\\
C_{ij}=i(i+n-2)+j(j+n-4)\label{eq 2340}
\eea
Now, from the condition\eqref{condition1} we find that only one representation of $O_R ^\delta (x)$ remains for each case in interest of $(\Delta,l)$ and the only possible $N_{a_1 a_2 a_3...a_l}^\Delta$ are
\begin{gather}
N^1_a=\pa_a O^0(x) \nonumber\\
N^1_{ab}=\left(\pa_a\pa_b-\frac{1}{n}\delta_{ab}\pa^2\right)O^{-1}(x)\nonumber\\
N^0_{ab}=\left(\pa_a O^{-1}_b+\pa_b O^{-1}_a-\frac{2}{n}\delta_{ab}\pa . O^{-1}\right)	
\end{gather}

\end{document}